\documentclass[
 reprint,
 amsmath,
 amssymb,
 aps,
]{revtex4-1}
\usepackage{graphicx}% Include figure files
\usepackage{dcolumn}% Align table columns on decimal point
\usepackage{bm}% bold math
\usepackage{glossaries}     % Enable the \newacronym and \gls commands to reference terms
\usepackage{commath}        % We have this here to use the \abs{} equation function
\usepackage{amssymb}
\usepackage{mathtools}
\usepackage{booktabs}
\usepackage{array}
\usepackage{makecell}
\usepackage{todonotes}

\usepackage{multirow}
\newcommand{\ang}{\mbox{\normalfont\AA}}

\newacronym{DFT}{DFT}{Density Functional Theory}
\newacronym{onlineal}{Online-AL}{Online Active Learning}
\newacronym{offlineal}{Offline-AL}{Offline Active Learning}

\newacronym{mlp}{MLP}{Machine Learning Potential}
\newacronym{QE}{QE}{Quantum Espresso}
\newacronym{VASP}{VASP}{Vienna Ab initio Simulation Package}
\newacronym{SI}{SI}{Supplementary Information}

\begin{document}

\title{Enabling robust offline active learning for machine learning potentials using simple physics-based priors}

\author{Muhammed Shuaibi, Saurabh Sivakumar, Rui Qi Chen}
%  \email{mshuaibi@andrew.cmu.edu}
\author{Zachary W. Ulissi}%
 \email{zulissi@andrew.cmu.edu}
\affiliation{%
Department of Chemical Engineering, Carnegie Mellon University, Pittsburgh, PA 15213, United States
}%

\date{\today}

\begin{abstract}
Machine learning surrogate models for quantum mechanical simulations has enabled the field to efficiently and accurately study material and molecular systems. Developed models typically rely on a substantial amount of data to make reliable predictions of the potential energy landscape or careful active learning and uncertainty estimates. When starting with small datasets, convergence of active learning approaches is a major outstanding challenge which limited most demonstrations to online active learning. In this work we demonstrate a $\Delta$-machine learning approach that enables stable convergence in offline active learning strategies by avoiding unphysical configurations.  We demonstrate our framework's capabilities on a structural relaxation, transition state calculation, and molecular dynamics simulation,  with the number of first principle calculations being cut down anywhere from 70-90\%.  The approach is incorporated and developed alongside AMP\textit{torch}, an open-source machine learning potential package, along with interactive Google Colab notebook examples.  
\end{abstract}

\maketitle

\section{Introduction}

The last decade has seen a surge in machine learning applications to material science, physics, and chemistry \cite{Artrith2014, Rupp2015, Natarajan2016, Peterson2016, Behler2016, Khorshidi2016, Bartok2010}. Characterizing a molecular system's potential energy surface (PES) has been a crucial step to the development of new catalysts and materials. Structure relaxation, molecular dynamics, and transition state calculations rely almost entirely on an accurate PES to serve their functions. \gls{mlp}s have demonstrated chemical accuracy at orders of magnitude faster computation times than traditional \textit{ab-initio} methods including density functional theory (DFT) and coupled cluster single double triple (CCSDT)  \cite{Zuo2020}. However, these demonstrations have generally required large datasets and careful uncertainty estimates. More importantly, the models developed have struggled to generalize to new systems and faced convergence issues when adding data, making the practicality of their day-to-day applications challenging \cite{Chen2020, Mueller2020, Behler2016, Schleder2019}. The potential of active learning in molecular simulations has not been fully realized due to convergence and implementation challenges. 

 The careful curation of training datasets for accurate molecular simulations has recently given way to active learning \cite{Vandermause2020, Jinnouchi2019, GarridoTorres2019, DelRio2019}. Active learning (AL) is the branch of machine learning concerned with systematically querying data points to be be part of the training set \cite{Settles2010}. The iterative process queries new data, trains a model, and repeats until a model performance is achieved. AL methods are particularly useful when the cost of querying data is substantial - as in the case of computing DFT. There are two main classes of strategies with relevance to molecular simulations. In \gls{onlineal}, configurations are generated sequentially using a \gls{mlp} and for each a decision is made whether to accept the estimate, perhaps using an uncertainty estimate. In \gls{offlineal}, a pool of candidates is generated and a decision is made which of the pool to add to the training set.

\begin{figure*}[!th]
    \centering
    \includegraphics[width=\textwidth]{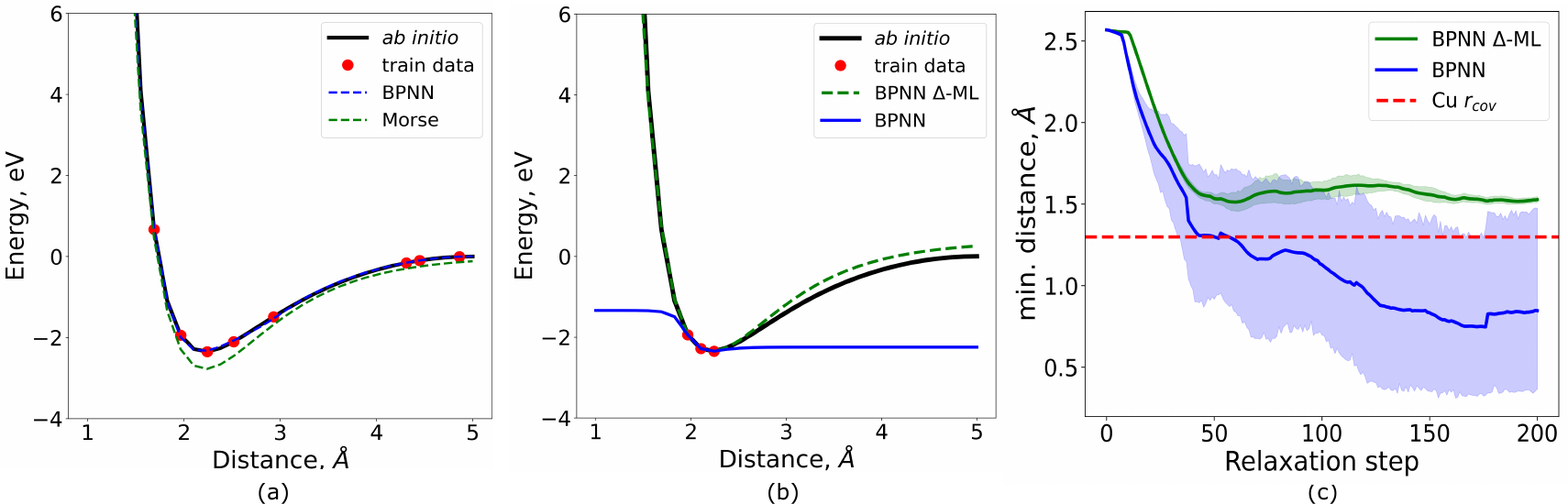}
    \caption{A traditional Behler-Parinello neural network (BPNN) trained to replicate the potential energy surface (PES) of a Cu-Cu bond with \textbf{(a)} a dataset spanning the PES and \textbf{(b)} a limited dataset trained with and without a Morse prior. \textbf{(c)} The minimum pair-wise distance of a structure relaxation carried out with a BPNN model, with and without the Morse prior. Relative to the covalent radius of Cu, our model consistently predicts more physically-consistent configurations as compared to the more unstable BPNN.}
    \label{fig:bond}
\end{figure*}  

Although there are many strategies available for both \gls{onlineal} and \gls{offlineal}, both commonly assume that all generated candidates are feasible to be queried and that adding data will not reduce accuracy on previous training data. Both of these assumptions are difficult with \gls{mlp}: \gls{DFT} often fails to converge on far-from-equilibrium structures, and many \gls{mlp} suffer if even a small number of configurations with large energies/forces are added to the training dataset \cite{GarridoTorres2019}. The most common approach to address these challenges is to carefully monitor uncertainty in the active learning process and prevent extrapolation to unphysical regions. This strategy is relatively straightforward to implement in \gls{onlineal}: if the uncertainty estimate is below a threshold, accept the prediction, otherwise run the DFT calculations. If the step size is small enough, the new configuration should be not so different from configurations in the training set. However, in \gls{offlineal} it is difficult to precisely define similarity between the pool of candidates and training set or predict which configurations are possible to converge with \gls{DFT}. Instead of solving this problem, we show that it is possible to mostly fix the underlying issues leading to unrealistic configurations.

In this work, we demonstrate that stable convergence in \gls{offlineal} with \gls{mlp} is possible by adding simple repulsive potentials and robust training procedures. This approach is implemented for the common combination of Behler-Parinello \gls{mlp} fingerprints with neural network atomic energy models \cite{Behler2007}. We show that a $\Delta$-ML approach with a base pairwise Morse potential and linear mixing rules is capable of sufficiently capturing the repulsive interactions between atoms that lead to \gls{DFT} errors. Since this Morse potential is not responsible for capturing the full potential, the parameterization only needs to be done once for each element. We demonstrate this approach for several types of calculations common in catalysis: structure relaxations, molecular dynamics, and transition state calculations. In each case, convergence with the addition of training data is essentially impossible with the base potential and well behaved with the $\Delta$-ML approach. In most cases this process allows for a reduction of 70-90\% in the number of DFT single-point evaluations necessary.  This process is further improved using standard neural network training approaches in the ML community to reduce the impact of random initial weights on small datasets. All of these are demonstrated in open-source and accessible AMP\textit{torch} GitHub repository with Google Colab ASE examples \cite{amptorch, examples}. 

\section{Methods}
The ML community continues to make advancements in the optimization and implementation of neural network based models \cite{Loshchilov2019, Fey2019, Paszke2019}. To leverage some of these approaches, we employ a Behler-Parinello neural network (BPNN). BPNNs construct element specific neural networks with the energy of the system the sum of atomic energy contributions. Per-atom forces are directly obtained from the negative gradient of the energy with respect to the atomic positions. We refer the readers to several reviews for a more detailed discussion on the BPNN model \cite{Behler2007, Behler2016, Khorshidi2016}. Additionally, neural network based models don't suffer the same kernel selection and scalability challenges that can come with Gaussian processes (GP) and other bayesian models \cite{Bartok2015}. Training neural networks, however, can be an extremely challenging task we hope to address in this work.

In the presence of an abundance of data, BPNN-like models have shown great success in replicating the PES of various systems \cite{Khorshidi2016, Peterson2016, Schran2020}. In the small data limit, however, neural network based models are unable to successfully characterize the energy surface, Figure \ref{fig:bond}b. More notably, model predictions are entirely ``physics-free", such that simple repulsive interactions are only ever learned by the model once enough data has been provided. As a result, a considerable amount of time may be wasted learning simple, widely understood, characteristics of the PES. Hybrid physics-based machine learning models can provide an important path forward to making reliable, physically-consistent discoveries in the sciences \cite{Willard, Karpatne2017}. To address this, we incorporate a $\Delta$-ML approach \cite{Ramakrishnan2014, Zhu2019} to learn the correction, $E_{NN}$, between a simple Morse potential, $\Delta E_{morse}$, and \textit{ab-initio} level theory - namely, DFT, $\Delta E_{DFT}(\textbf{x})$:

\begin{flalign}
    \Delta E_{DFT}(\textbf{x}) &= E_{DFT}(\textbf{x}) - E_{DFT}(x_{ref})&&\\
    \Delta E_{morse}(\textbf{x}) &= E_{morse}(\textbf{x}) - E_{morse}(x_{ref})&&\\
    E_{NN}(\textbf{x}) &= \Delta E_{DFT}(\textbf{x}) - \Delta E_{morse}(\textbf{x})&&\\
    E_{morse+NN}(\textbf{x}) &= \Delta E_{morse}(\textbf{x}) + E_{DFT}(x_{ref}) \\\nonumber&\hspace{3.5cm}+E_{NN}(\textbf{x})
\end{flalign}
Where $E_{DFT}(x_{ref})$ and $E_{morse}(x_{ref})$ correspond to reference energies necessary to correct for differences in their absolute energies. Reference energies are computed from a same arbitrary structure, $x_{ref}$; the dataset's first structure was used in our applications. Per-element parameters of the Morse potential, $D_e$, $r_e$, and $a$, are fitted to DFT data \textit{a priori}. A more detailed description of the fitting procedure is included in the \gls{SI}. By leveraging the Morse potential as the backbone to the model, the ML component is allowed to learn the remaining functional form while still capturing physics-based repulsive interactions previously missed. Additionally, learning a correction can allow the neural network to learn a much smoother function than the underlying PES, improving training stability and convergence.

\begin{figure*}[ht!]
    \centering
    \includegraphics[width=0.48\textwidth]{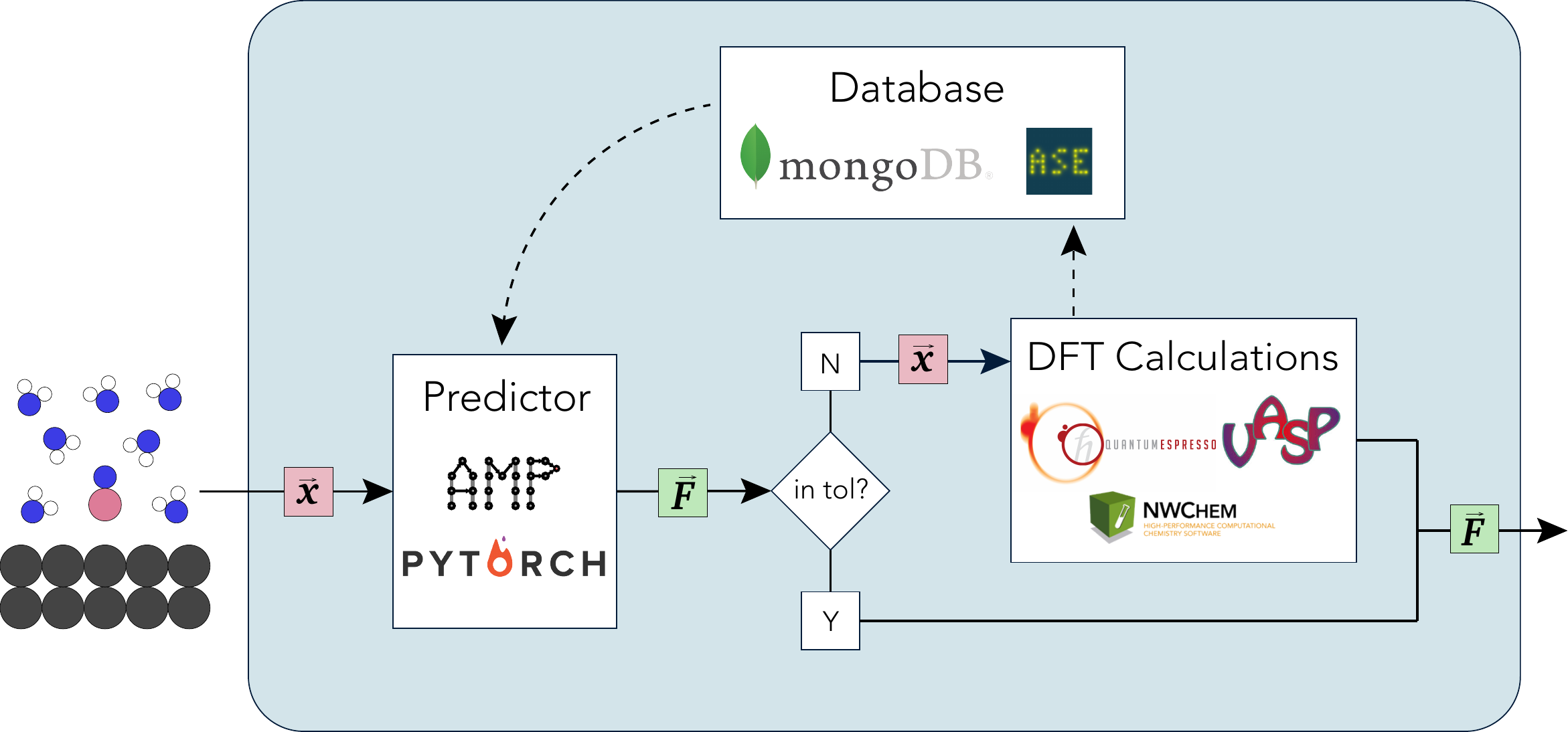}
    \includegraphics[width=0.48\textwidth]{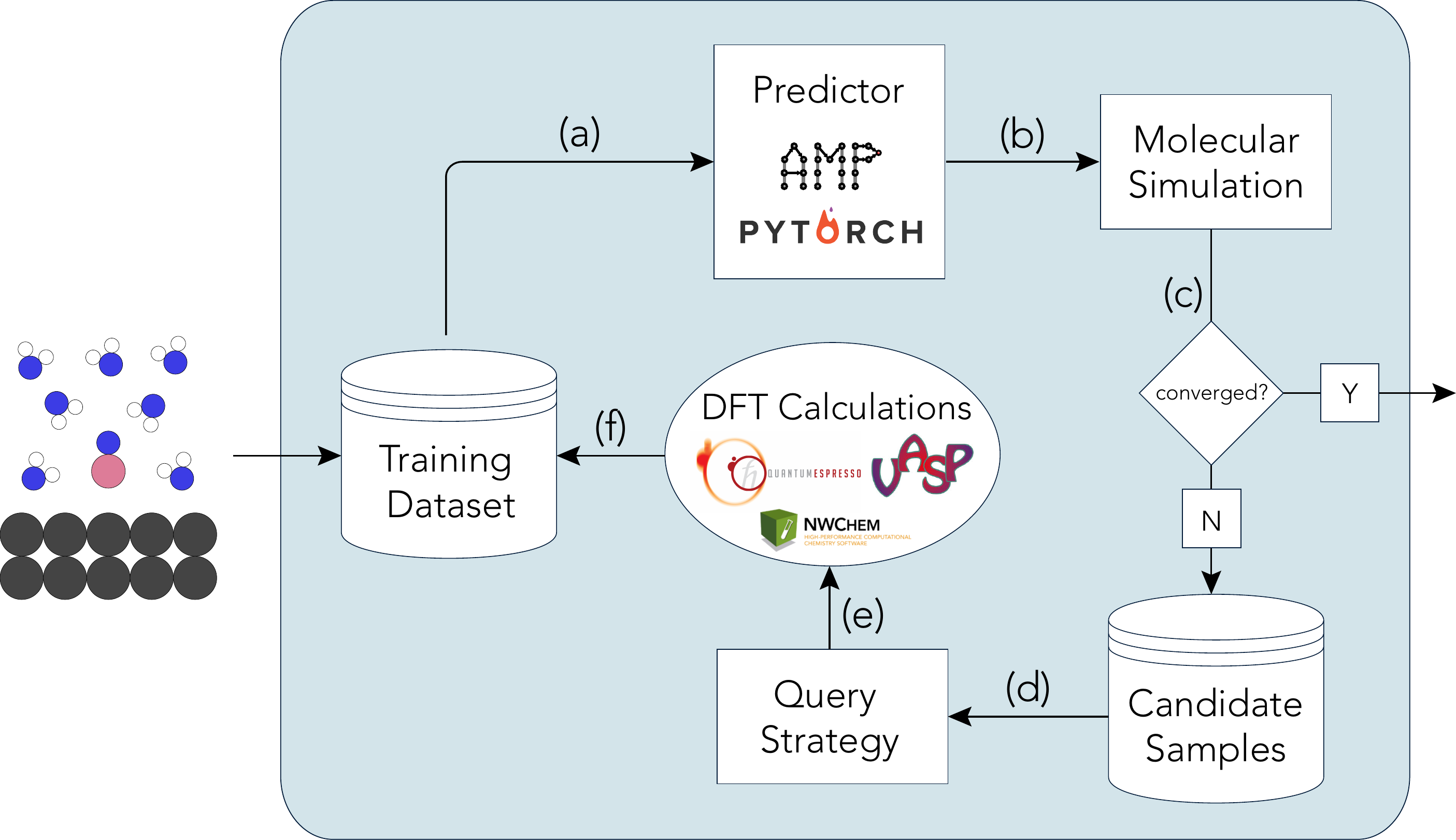}
    \caption{Online and Offline active learning frameworks to accelerate molecular simulations.
    \textbf{Left}: \gls{onlineal}. At each time step, our ML model makes a prediction of the energy and forces and assesses the uncertainty of its estimate. If confident, the ML results are used to take a step in the molecular simulation. Otherwise, a DFT call is made, added to a database, and the model retrained. \textbf{Right}: Proposed \gls{offlineal}.
    \textbf{(a)} An initial training dataset is used to train the ML model; \textbf{(b)} the trained ML model runs the atomistic simulation of interest; \textbf{(c)} termination if converged, otherwise, the generated data is stored as a pool of potential candidates; \textbf{(d)} a query strategy identifies what points to be added to the training set; \textbf{(e)} \textit{ab-initio} calculations are performed on selected candidates; \textbf{(f)} queried points are added to the training set. Repeat until convergence is reached}
    \label{fig:framework}
\end{figure*}

We illustrate the benefits of this simple Morse potential by running a structure relaxation of carbon on copper (C/Cu) with our model trained on a single image (Figure \ref{fig:bond}c). The minimum pair-wise distance of the resulting trajectory are compared to that not employing a morse potential. Our model consistently predicts configurations above the covalent radius of copper, a good indication repulsive forces are being captured. On the other hand, a traditional BPNN shows wide variations while on average predicting configurations well below the more stable covalent radius.

The fitting of MLPs is an important process in our AL framework, as they are responsible for generating candidates for training data. A poorly fit MLP may generate unfeasible candidates that DFT can not converge on. This is especially true when working without a physics-based potential. Working within the small data regime allows us to leverage quasi-newton optimizers, namely LBFGS. LBFGS and other second order optimizers provides us with improved convergence of our model training over standard first order methods such as SGD and Adam. This advantage, however, is only really feasible in the small data limit where the computational cost of such methods can be afforded. Additionally, we incorporate a cosine annealing learning rate scheduler with warm restarts \cite{loshchilov2016sgdr} to aid in the convergence of the \gls{offlineal} framework. A more detailed comparison  can be found in the \gls{SI}.

Similar to previous works \cite{Vandermause2020, Jinnouchi2019, Jinnouchi2019}, our \gls{onlineal} framework begins with little to no data and must identify the right points to query and improve the model over the course of a molecular simulation (Figure \ref{fig:framework}). Rather than relying on kernel-based models, our \gls{onlineal} framework utilizes the proposed physics-coupled BPNN. We incorporate bootstrap-ensembling, or bagging, in order to quantify our model's uncertainty. Bagging involves training multiple, randomly initialized, independent models with training sets randomly sampled, \textit{with} replacement, from an original dataset \cite{Peterson2017}. Predictions and uncertainty estimations are then calculated from the ensemble statistics.

An offline active learning can offer model and computational advantages over \gls{onlineal} frameworks. Rather than making query decisions in a dynamic process, we present a method to select from a pool of candidates. We accomplish this by iteratively running an ML-driven molecular simulation. After each iteration, a querying strategy samples from the generated trajectory. Queried points are then evaluated with DFT, added to an original dataset, and the ML model retrained (Figure \ref{fig:framework}). The process is repeated until a defined convergence criteria is met. Despite the ML model resulting in inaccurate simulations early on, diverse, informative configurations are generated to train the ML model. In dealing with a pool of query candidates, the framework allows us to explore more sophisticated querying strategies rather than being limited to strictly uncertainty estimates of \gls{onlineal} \cite{Settles2010}. Additionally, the reliance on uncertainty estimates can pose more fundamental questions surrounding how trustworthy a model's estimates really are \cite{Tran2020}. 

We demonstrate the proposed framework on several common catalysis applications: structure relaxations, transition-state calculations, and molecular dynamics. A random sample query strategy is introduced in the \gls{offlineal} schemes to demonstrate the effectiveness of even the simplest of query strategies over \gls{onlineal}. More problem-specific query strategies are proposed for structural relaxations and transition-state calculations, further improving the convergence. To show the generality of this approach in small-data applications, we also use two common \gls{DFT} packages - \gls{VASP} and \gls{QE} \cite{Kresse1993,Kresse1996, Giannozzi2009}. The use of \gls{QE} allows for interactive and open demonstrations of this approach.  Several Google Colab notebooks have been included in the \gls{SI} allowing users to easily experiment and explore new systems with AMP\textit{torch} and \gls{QE} without needing to locally install and manage dependencies.

%%%%% Relaxation %%%%%

\section{Results and discussion}
A structural relaxation is performed for C/Cu(100) with cell size $2\times2\times3$. An initial guess of $3\ang$ from the surface is made for the adsorbate. Periodic boundary conditions are applied in the x and y directions and the last slab layer is fixed from relaxations. 

Performance is measured by the final structure and energy mean-absolute-errors (MAE). A random sample query strategy selects configurations from the generated relaxations to be queried. We run the \gls{offlineal} framework under a variety of batching scenarios, terminating after \textit{N} iterations, sampling \textit{M} configurations per iteration, for an arbitrary total of $NM = 20$ DFT calls. Results are summarized in Table \ref{table:batch}.

Under the above random query strategy, systematic termination of the \gls{offlineal} loop is quite heuristic. To address this, we incorporate an alternative query and termination strategy. At each iteration, in addition to a random configuration, the predicted relaxed structure is also queried. If the predicted relaxed structure's max per-atom force, as evaluated by DFT, is below the optimizer's convergence criteria, the AL loop is terminated. Otherwise, the configurations are added to the original dataset, and the framework cycles. In querying the model's predicted relaxed structure we are assured in our framework's ability to accurately reach a local minima. 

We compare the performance of this \gls{offlineal} scheme and \gls{onlineal} with and without the $\Delta$-ML in Table \ref{table:al_table}. \gls{offlineal} and \gls{onlineal} tolerances correspond to the max per-atom force termination criteria and max force variance tolerated by the ensemble, respectively. Force termination criteria of 0.03 and 0.05 eV/$\ang$ are compared to explore the tradeoff between accuracy and number of DFT calls. \gls{onlineal} was empirically set to query a DFT call when the ensemble based force uncertainty reached above a threshold of 0.05 eV/$\ang$. The energy and structure MAE associated with the system's initial structure is 2.82 eV and 0.15 \ang, respectively. Our best performing framework - \gls{offlineal} with $\Delta$-ML (0.03 eV/$\ang$), reported average energy and structure MAEs of 0.0039 eV and 0.0032 $\ang$ with 17 total DFT calls - a $66.7\%$ reduction. Without the inclusion of the Morse prior, a standard BPNN was unable to converge, generating configuration that DFT was unable to evaluate in almost all our experiments.

\begin{table}[t]
\centering
\begin{tabular}[c]{|>{\centering\arraybackslash}m{1.8cm}|>{\centering\arraybackslash}m{2cm}|>{\centering\arraybackslash}m{1.8cm}|>{\centering\arraybackslash}m{2cm}|>{\centering\arraybackslash}m{1.2cm}|}
\hline
\multicolumn{2}{|c|}{Batching Scenario}&& \\
Iterations & Samples per iteration& Energy MAE (eV) & Structure MAE (\ang)\\
\hline\hline
20 & 1 & 0.0063 & 0.0037 \\ \hline
10 & 2 & 0.0069 & 0.0063 \\ \hline
5 & 4 & 0.0080 & 0.0067 \\ \hline
\end{tabular}
\caption{Comparison of different offline active learning batching scenarios on the structural relaxation of C/Cu(100). At each iteration, a varying number of queries are randomly made from the generated relaxation.  A tradeoff in performance and the number of samples per iteration is observed for a fixed total number of DFT calls $= 20$. All models trained here incorporated the proposed Morse prior.}
\label{table:batch}
\end{table}

\begin{table}[t]
\centering
\begin{tabular}[c]{|>{\centering\arraybackslash}m{2cm}|>{\centering\arraybackslash}m{1.4cm}|>{\centering\arraybackslash}m{1.6cm}|>{\centering\arraybackslash}m{1.6cm}|>{\centering\arraybackslash}m{1.2cm}|}
\hline
Framework (tolerance) & MLP & Energy MAE (eV) & Structure MAE (\ang) & DFT calls\\
\hline\hline
 \break DFT & - & - & - & 51\\ \hline
\gls{offlineal} (0.03 eV/$\ang$) & BPNN $\Delta$-ML & 0.0039 & 0.0032 & 17\\ \hline
\gls{offlineal} (0.05 eV/$\ang$) & BPNN $\Delta$-ML & 0.0049 & 0.0059 & 15\\ \hline
\gls{offlineal} (0.05 eV/$\ang$) & BPNN only & \multicolumn{3}{c|}{does not converge}\\ \hline
\gls{onlineal} (0.05 eV/$\ang$) & BPNN $\Delta$-ML & 0.0073 & 0.0107 & 30\\ \hline
\gls{onlineal} (0.05 eV/$\ang$) & BPNN only & 0.2884 & 0.0263 & 22\\ \hline
\end{tabular}
\caption{Summary of the various strategies' performance on the structural relaxation of C/Cu(100). The effects of the Morse prior on the convergence of both the offline and online active learning are also shown. The querying strategy employed by the \gls{offlineal} framework relies on a quasi-random strategy, additionally sampling and assessing convergence on the framework's generated relaxed structure.}
\label{table:al_table}
\end{table}

\begin{figure*}[ht]
    \centering
    \includegraphics[width=\textwidth]{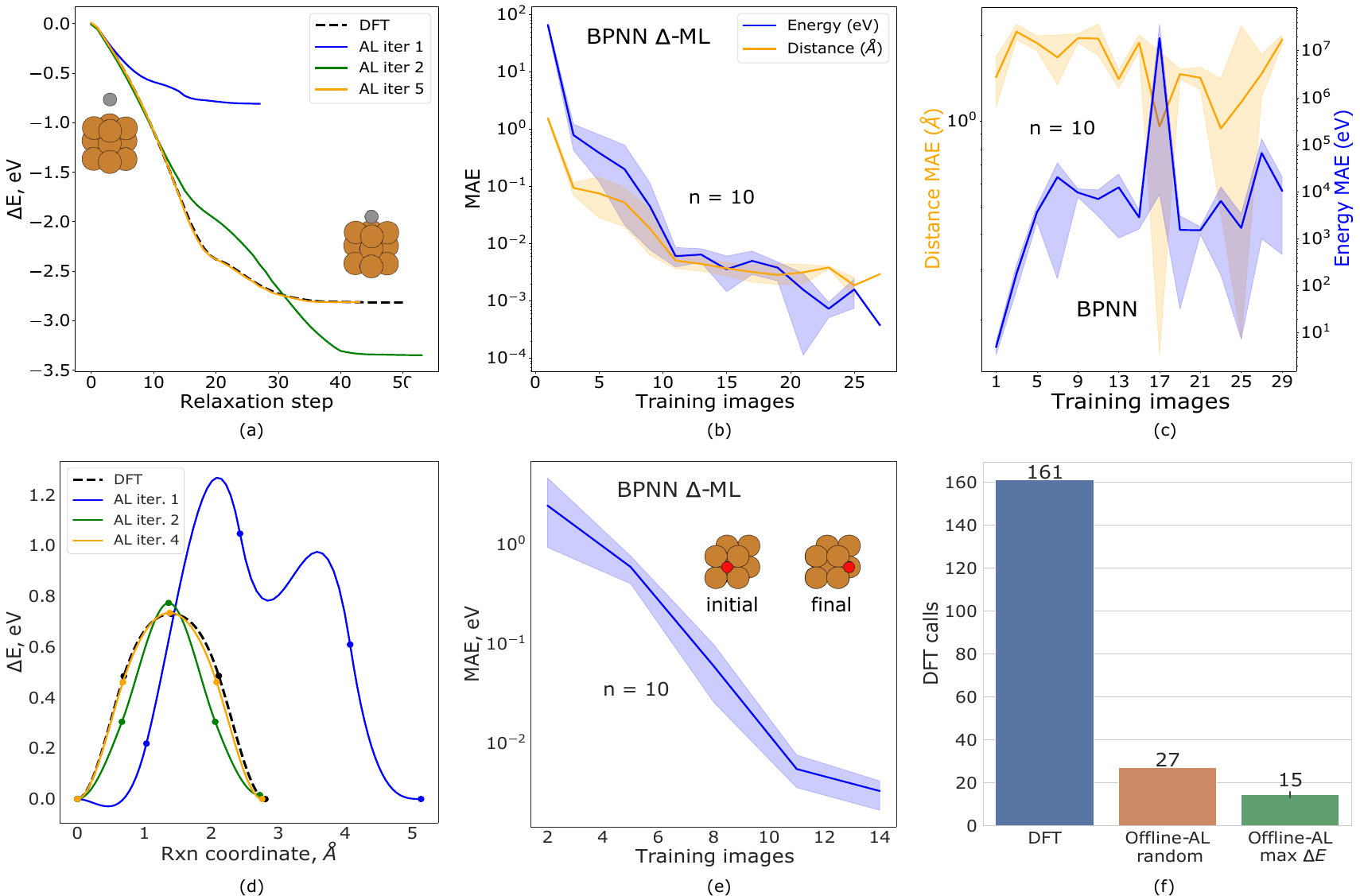}
    \caption{ \gls{offlineal} applications to structural relaxations and transition state calculations. \textbf{(a)} Evolution of the structural relaxation of C on Cu(100) over a few cycles of the \gls{offlineal} \textbf{(b)} Relaxed structure and energy learning curves of the \gls{offlineal} framework, using the BPNN $\Delta$-ML model. \textbf{(c)} Convergence instability associated with not incorporating the Morse potential prior in an \gls{offlineal} context. \textbf{(d)} Evolution of the transition state calculation for the surface diffusion of O on Cu(100).  \textbf{(e)} Learning curve associated with the energy barrier of the oxygen diffusion example of (d). \textbf{(f)} Total number of DFT calls queried by the \gls{offlineal} under different querying strategies for the energy barrier associated with the diffusion of oxygen on copper.
    }
    
    \label{fig:al_relax}
\end{figure*}

%%%%% NEB %%%%%

Next, we demonstrate an application to transition state calculations, specifically, nudge-elastic-band (NEB) methods \cite{Henkelman2000, Henkelman2000_2}. NEB calculations require defining the initial and final structures for the transition state to be calculated. Machine-learning accelerated NEB calculations have typically relied on \textit{ab-initio} relaxed initial and final structures, a costly step of a NEB calculation \cite{Ang2020}. In fixing the initial and final structures, the machine learning objective is simplified to an interpolation problem. We demonstrate our framework's ability to accelerate the complete NEB calculation, including initial and final structure relaxations, to find the surface diffusion energy barrier of oxygen on Cu(100). To illustrate our framework, we use five images to build the NEB including the initial and final states which have not been relaxed previously. 

The convergence evolution of our \gls{offlineal} framework is illustrated in Figure \ref{fig:al_relax}d, approaching the true energy barrier after a few iterations. Similarly, convergence was not achieved, with often failing DFT evaluations, without the inclusion of the Morse prior. In addition to a random query strategy, we compare the impacts a more crafted query strategy can make on the number of DFT calls (Figure \ref{fig:al_relax}e). Two strategies are compared. The first, a simple random strategy where images used to build the NEB are randomly sampled from generated NEBs and evaluated using DFT before being added to the training data. The second, a strategy tailor-made for the NEBs where the highest energy point and initial and final points are sampled at each iteration. The loop is terminated once the difference between the ML predicted energy and DFT evaluated energy of the ML predicted saddle point meets a specified threshold. Both cases demonstrate a significant reduction in the number of DFT calls required to construct the NEB. 

% MD
\begin{figure*}
    \centering
    \includegraphics[width=\textwidth]{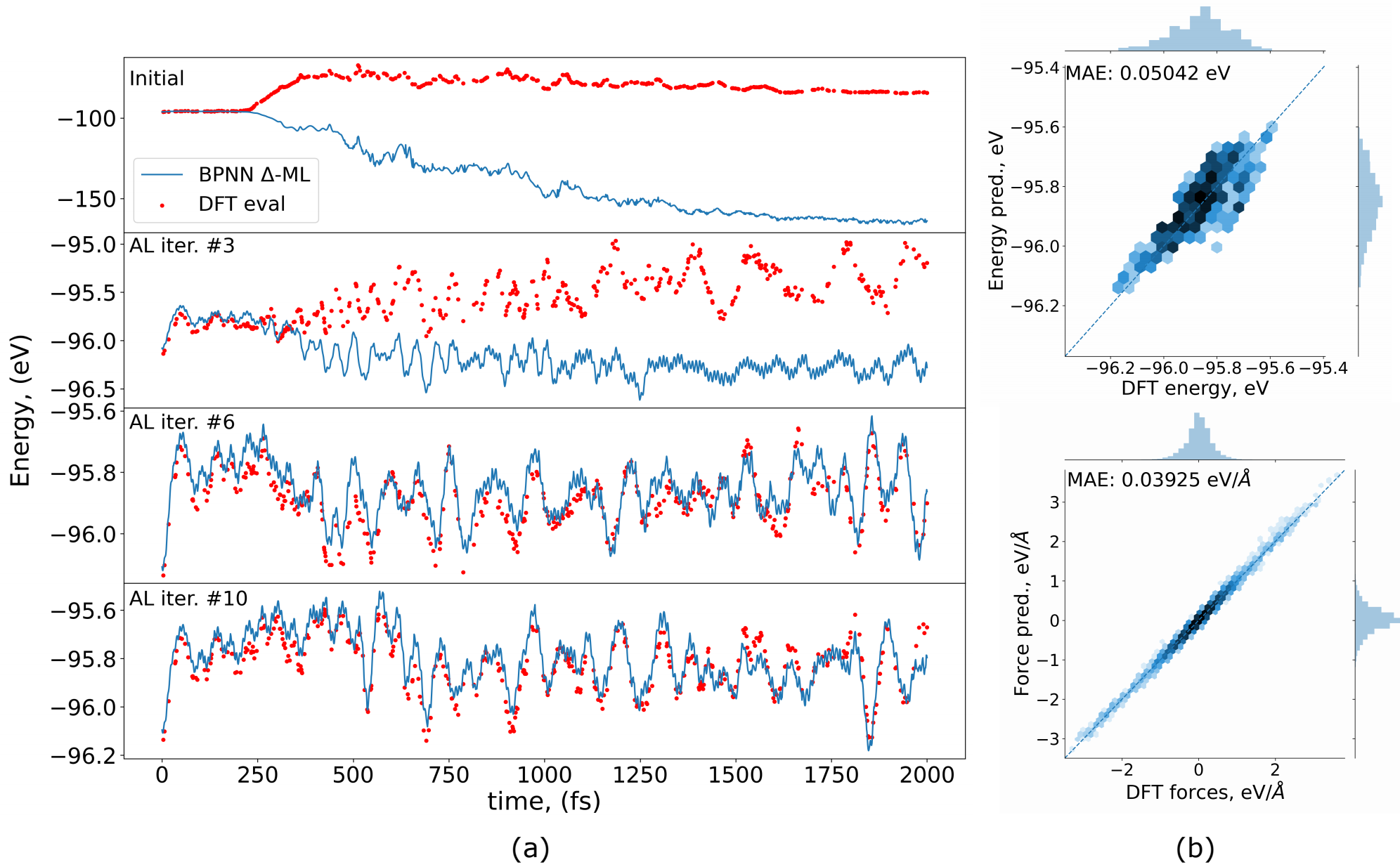}
    \caption{ \gls{offlineal} demonstration to a 2ps MD simulation of CO on Cu(100) \textbf{(a)} Evolution of the MD trajectory over several iterations of the active learning framework. We verify the effectiveness of our framework by randomly sampling configurations and comparing DFT evaluated energy and forces with that of our model's predictions. \textbf{(b)} Parity plots associated with the DFT evaluated configurations and our model's predictions, demonstrating good agreement.
    }
    \label{fig:al_md}
\end{figure*}

\begin{figure}
    \centering
    \includegraphics[width=0.5\textwidth]{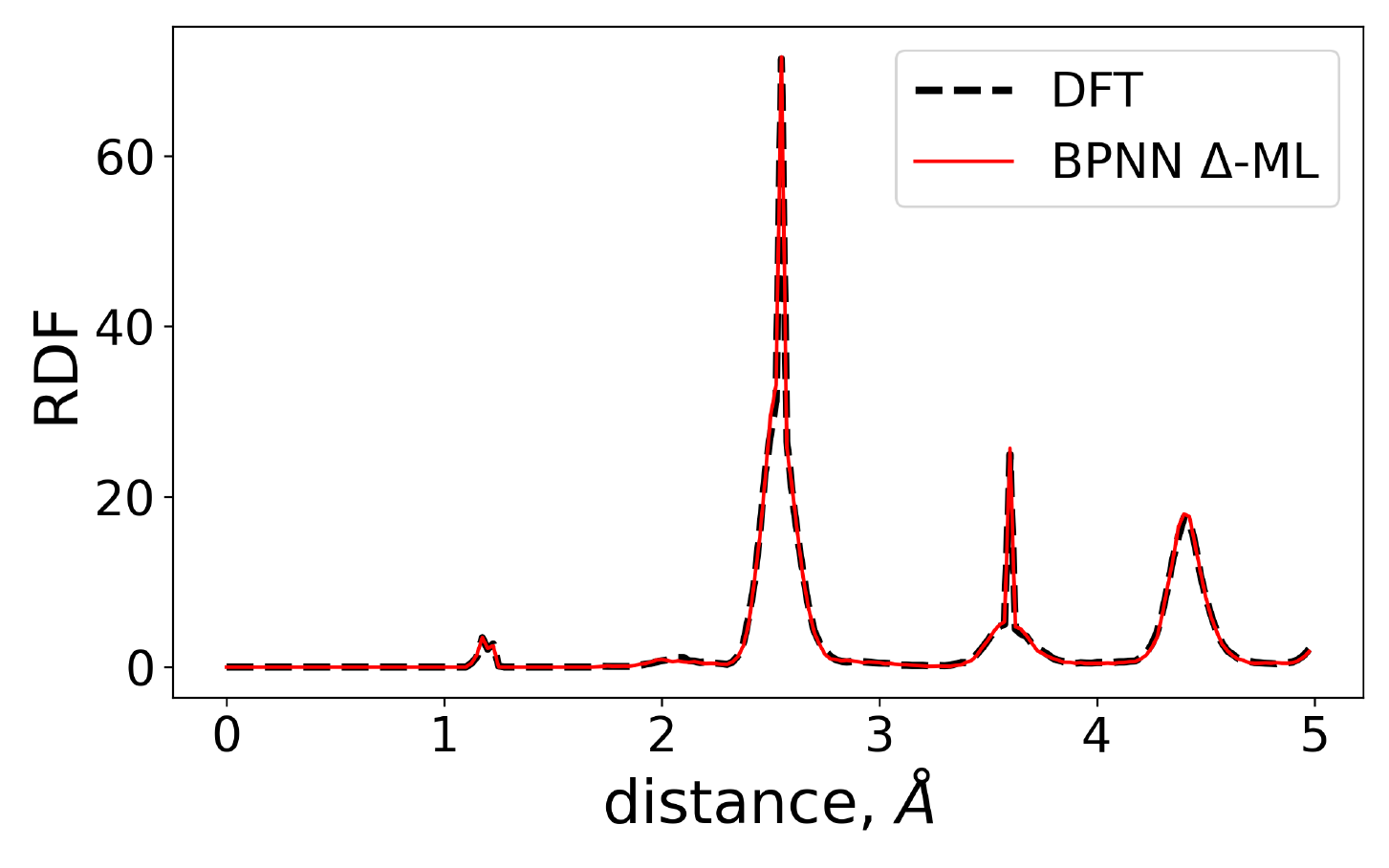}
    \caption{Radial distribution function (RDF) of the ground truth DFT and our framework's 6th iteration for the MD simulation of CO/Cu(100). Demonstrating good consistency even before the allotted number of iterations.}
    \label{fig:al_rdf}
\end{figure}

Machine learning surrogates to DFT are considerably favorable in the context of long time-scale simulations, namely, molecular dynamics (MD). Unlike structural relaxations, MD simulations are typically carried out on orders of magnitudes more steps. Several works have addressed these challenges through GP-based \gls{onlineal} frameworks \cite{Vandermause2020, Jinnouchi2019}. We demonstrate that our proposed \gls{offlineal} framework is even capable of converging to an accurate MD simulation with an initialized dataset of size 1. A 2ps MD simulation of CO on Cu(100) in a 300K NVT ensemble is used for our demonstration. 

 Beginning with a single data point, our framework cycled for 10 iterations, randomly querying 50 configurations for a total of 500 DFT calls by the end of our experiment. Unlike structural relaxations with a well defined target, MD simulations are more stochastic in nature and are unlikely to follow an identical trajectory over multiple iterations. To demonstrate the effectiveness of our framework, we verify the performance, at each iteration, by randomly sampling 400 configurations from our ML predicted trajectory and validate their corresponding energy and force predictions with DFT. We illustrate the iterative convergence of our framework in Figure \ref{fig:al_md}. Despite the upper limit of 10 iterations, we observe good agreement with DFT by iteration 6 - a reduction of 85\% in DFT calls. Additionally, we demonstrate consistency in the radial distribution function of our framework's generated simulation to that of the original DFT simulation (Figure \ref{fig:al_rdf}).

\section{Conclusion}

The development of accurate and reliable \gls{mlp} has been a challenging task for the community. The careful curation of datasets is especially difficult in trying to generalize to new systems. Active learning has provided promising results in accelerating molecular simulations while minimizing risks of extrapolation. Neural-network based models, however, have struggled with such demonstrations for their reliance on large amounts of data. As deep learning research continues to make significant strides, understanding how to better incorporate neural network based \gls{mlp}s into active learning pipelines can help provide more accurate and robust models.

This Letter presented a neural-network based offline active-learning framework to accelerate a variety of molecular simulations beginning with extremely limited data. We introduced a physics-based prior, Morse potential, into our model in a $\Delta$-ML manner, to capture basic repulsive interactions crucial in the convergence of our framework. We demonstrate the framework's ability to accurately converge simulations including structural relaxations, molecular dynamics simulations, and transition-state calculations. In each of these, the proportion of DFT calls reduced were 71\%, 75\%, and 91\%, respectively. The framework presented is extremely flexible, allowing users to define their own querying strategies, termination criteria, and incorporate their own, more complex molecular simulations they wish to accelerate with AMP\textit{torch}. Future directions will explore more systematic querying strategies and termination criteria to further accelerate the framework while being robust to larger, more complex systems. Additionally, exploring alternative model priors can help improve the performance and generalizability of the overall framework. 

\section{Calculation Settings}

Single-point \gls{DFT} calculations were performed \textit{Quantum Espresso (QE)\cite{Giannozzi2009}} implemented in \textit{ASE} \cite{HjorthLarsen2017}; using the \textit{PBE} exchange-correlation functional \cite{Perdew1996}; a plane wave basis set with an energy-cutoff of 500 eV; k-points of $4 \times 4 \times 1$; and the pseudopotentials provided by Garrity, et al.\cite{Garrity2014}. The same settings were also used for \gls{DFT} calculations in fitting the Morse potential parameters. The following tools and settings were used for our DFT calculations: \textit{\gls{VASP} 5.4.4.18} \cite{Kresse1993, Kresse1996}; using the \textit{PBE} exchange-correlation functional; a plane wave basis set with an energy-cutoff of 400eV; and k-points of $4\times4\times1$. AMP\textit{torch} \cite{amptorch} was used for all machine learning and active learning components of the framework.

\section{Acknowledgements}

We acknowledge the support from the U.S. Department of Energy, Office of Science, Basic Energy Sciences Award \#{}DE-FOA-0001912. Additionally, we thank Andrew Peterson and AJ Medford for their thoughtful discussions.

\bibliography{main_paper}% Produces the bibliography via BibTeX.

\begin{thebibliography}{40}%
\makeatletter
\providecommand \@ifxundefined [1]{%
 \@ifx{#1\undefined}
}%
\providecommand \@ifnum [1]{%
 \ifnum #1\expandafter \@firstoftwo
 \else \expandafter \@secondoftwo
 \fi
}%
\providecommand \@ifx [1]{%
 \ifx #1\expandafter \@firstoftwo
 \else \expandafter \@secondoftwo
 \fi
}%
\providecommand \natexlab [1]{#1}%
\providecommand \enquote  [1]{``#1''}%
\providecommand \bibnamefont  [1]{#1}%
\providecommand \bibfnamefont [1]{#1}%
\providecommand \citenamefont [1]{#1}%
\providecommand \href@noop [0]{\@secondoftwo}%
\providecommand \href [0]{\begingroup \@sanitize@url \@href}%
\providecommand \@href[1]{\@@startlink{#1}\@@href}%
\providecommand \@@href[1]{\endgroup#1\@@endlink}%
\providecommand \@sanitize@url [0]{\catcode `\\12\catcode `\$12\catcode
  `\&12\catcode `\#12\catcode `\^12\catcode `\_12\catcode `\%12\relax}%
\providecommand \@@startlink[1]{}%
\providecommand \@@endlink[0]{}%
\providecommand \url  [0]{\begingroup\@sanitize@url \@url }%
\providecommand \@url [1]{\endgroup\@href {#1}{\urlprefix }}%
\providecommand \urlprefix  [0]{URL }%
\providecommand \Eprint [0]{\href }%
\providecommand \doibase [0]{http://dx.doi.org/}%
\providecommand \selectlanguage [0]{\@gobble}%
\providecommand \bibinfo  [0]{\@secondoftwo}%
\providecommand \bibfield  [0]{\@secondoftwo}%
\providecommand \translation [1]{[#1]}%
\providecommand \BibitemOpen [0]{}%
\providecommand \bibitemStop [0]{}%
\providecommand \bibitemNoStop [0]{.\EOS\space}%
\providecommand \EOS [0]{\spacefactor3000\relax}%
\providecommand \BibitemShut  [1]{\csname bibitem#1\endcsname}%
\let\auto@bib@innerbib\@empty
%</preamble>
\bibitem [{\citenamefont {Artrith}\ and\ \citenamefont
  {Kolpak}(2014)}]{Artrith2014}%
  \BibitemOpen
  \bibfield  {author} {\bibinfo {author} {\bibfnamefont {N.}~\bibnamefont
  {Artrith}}\ and\ \bibinfo {author} {\bibfnamefont {A.~M.}\ \bibnamefont
  {Kolpak}},\ }\href {\doibase 10.1021/nl5005674} {\bibfield  {journal}
  {\bibinfo  {journal} {Nano Letters}\ } (\bibinfo {year} {2014}),\
  10.1021/nl5005674}\BibitemShut {NoStop}%
\bibitem [{\citenamefont {Rupp}\ \emph {et~al.}(2015)\citenamefont {Rupp},
  \citenamefont {Ramakrishnan},\ and\ \citenamefont {von
  Lilienfeld}}]{Rupp2015}%
  \BibitemOpen
  \bibfield  {author} {\bibinfo {author} {\bibfnamefont {M.}~\bibnamefont
  {Rupp}}, \bibinfo {author} {\bibfnamefont {R.}~\bibnamefont {Ramakrishnan}},
  \ and\ \bibinfo {author} {\bibfnamefont {O.~A.}\ \bibnamefont {von
  Lilienfeld}},\ }\href@noop {} {\bibfield  {journal} {\bibinfo  {journal}
  {Journal of Physical Chemistry Letters}\ } (\bibinfo {year}
  {2015})}\BibitemShut {NoStop}%
\bibitem [{\citenamefont {Natarajan}\ and\ \citenamefont
  {Behler}(2016)}]{Natarajan2016}%
  \BibitemOpen
  \bibfield  {author} {\bibinfo {author} {\bibfnamefont {S.~K.}\ \bibnamefont
  {Natarajan}}\ and\ \bibinfo {author} {\bibfnamefont {J.}~\bibnamefont
  {Behler}},\ }\href {\doibase 10.1039/c6cp05711j} {\bibfield  {journal}
  {\bibinfo  {journal} {Physical Chemistry Chemical Physics}\ } (\bibinfo
  {year} {2016}),\ 10.1039/c6cp05711j}\BibitemShut {NoStop}%
\bibitem [{\citenamefont {Peterson}(2016)}]{Peterson2016}%
  \BibitemOpen
  \bibfield  {author} {\bibinfo {author} {\bibfnamefont {A.~A.}\ \bibnamefont
  {Peterson}},\ }\href {\doibase 10.1063/1.4960708} {\bibfield  {journal}
  {\bibinfo  {journal} {Journal of Chemical Physics}\ } (\bibinfo {year}
  {2016}),\ 10.1063/1.4960708}\BibitemShut {NoStop}%
\bibitem [{\citenamefont {Behler}(2016)}]{Behler2016}%
  \BibitemOpen
  \bibfield  {author} {\bibinfo {author} {\bibfnamefont {J.}~\bibnamefont
  {Behler}},\ }\href {\doibase 10.1063/1.4966192} {\bibfield  {journal}
  {\bibinfo  {journal} {Journal of Chemical Physics}\ } (\bibinfo {year}
  {2016}),\ 10.1063/1.4966192}\BibitemShut {NoStop}%
\bibitem [{\citenamefont {Khorshidi}\ and\ \citenamefont
  {Peterson}(2016)}]{Khorshidi2016}%
  \BibitemOpen
  \bibfield  {author} {\bibinfo {author} {\bibfnamefont {A.}~\bibnamefont
  {Khorshidi}}\ and\ \bibinfo {author} {\bibfnamefont {A.~A.}\ \bibnamefont
  {Peterson}},\ }\href {\doibase 10.1016/j.cpc.2016.05.010} {\bibfield
  {journal} {\bibinfo  {journal} {Computer Physics Communications}\ } (\bibinfo
  {year} {2016}),\ 10.1016/j.cpc.2016.05.010}\BibitemShut {NoStop}%
\bibitem [{\citenamefont {Bart{\'{o}}k}\ \emph {et~al.}(2010)\citenamefont
  {Bart{\'{o}}k}, \citenamefont {Payne}, \citenamefont {Kondor},\ and\
  \citenamefont {Cs{\'{a}}nyi}}]{Bartok2010}%
  \BibitemOpen
  \bibfield  {author} {\bibinfo {author} {\bibfnamefont {A.~P.}\ \bibnamefont
  {Bart{\'{o}}k}}, \bibinfo {author} {\bibfnamefont {M.~C.}\ \bibnamefont
  {Payne}}, \bibinfo {author} {\bibfnamefont {R.}~\bibnamefont {Kondor}}, \
  and\ \bibinfo {author} {\bibfnamefont {G.}~\bibnamefont {Cs{\'{a}}nyi}},\
  }\href {\doibase 10.1103/PhysRevLett.104.136403} {\bibfield  {journal}
  {\bibinfo  {journal} {Physical Review Letters}\ } (\bibinfo {year} {2010}),\
  10.1103/PhysRevLett.104.136403},\ \Eprint {http://arxiv.org/abs/0910.1019}
  {arXiv:0910.1019} \BibitemShut {NoStop}%
\bibitem [{\citenamefont {Zuo}\ \emph {et~al.}(2020)\citenamefont {Zuo},
  \citenamefont {Chen}, \citenamefont {Li}, \citenamefont {Deng}, \citenamefont
  {Chen}, \citenamefont {Behler}, \citenamefont {Cs{\'{a}}nyi}, \citenamefont
  {Shapeev}, \citenamefont {Thompson}, \citenamefont {Wood},\ and\
  \citenamefont {Ong}}]{Zuo2020}%
  \BibitemOpen
  \bibfield  {author} {\bibinfo {author} {\bibfnamefont {Y.}~\bibnamefont
  {Zuo}}, \bibinfo {author} {\bibfnamefont {C.}~\bibnamefont {Chen}}, \bibinfo
  {author} {\bibfnamefont {X.}~\bibnamefont {Li}}, \bibinfo {author}
  {\bibfnamefont {Z.}~\bibnamefont {Deng}}, \bibinfo {author} {\bibfnamefont
  {Y.}~\bibnamefont {Chen}}, \bibinfo {author} {\bibfnamefont {J.}~\bibnamefont
  {Behler}}, \bibinfo {author} {\bibfnamefont {G.}~\bibnamefont
  {Cs{\'{a}}nyi}}, \bibinfo {author} {\bibfnamefont {A.~V.}\ \bibnamefont
  {Shapeev}}, \bibinfo {author} {\bibfnamefont {A.~P.}\ \bibnamefont
  {Thompson}}, \bibinfo {author} {\bibfnamefont {M.~A.}\ \bibnamefont {Wood}},
  \ and\ \bibinfo {author} {\bibfnamefont {S.~P.}\ \bibnamefont {Ong}},\ }\href
  {\doibase 10.1021/acs.jpca.9b08723} {\bibfield  {journal} {\bibinfo
  {journal} {Journal of Physical Chemistry A}\ } (\bibinfo {year} {2020}),\
  10.1021/acs.jpca.9b08723},\ \Eprint {http://arxiv.org/abs/1906.08888}
  {arXiv:1906.08888} \BibitemShut {NoStop}%
\bibitem [{\citenamefont {Chen}\ \emph {et~al.}(2020)\citenamefont {Chen},
  \citenamefont {Zuo}, \citenamefont {Ye}, \citenamefont {Li}, \citenamefont
  {Deng},\ and\ \citenamefont {Ong}}]{Chen2020}%
  \BibitemOpen
  \bibfield  {author} {\bibinfo {author} {\bibfnamefont {C.}~\bibnamefont
  {Chen}}, \bibinfo {author} {\bibfnamefont {Y.}~\bibnamefont {Zuo}}, \bibinfo
  {author} {\bibfnamefont {W.}~\bibnamefont {Ye}}, \bibinfo {author}
  {\bibfnamefont {X.}~\bibnamefont {Li}}, \bibinfo {author} {\bibfnamefont
  {Z.}~\bibnamefont {Deng}}, \ and\ \bibinfo {author} {\bibfnamefont {S.~P.}\
  \bibnamefont {Ong}},\ }\href {\doibase 10.1002/aenm.201903242} {\bibfield
  {journal} {\bibinfo  {journal} {Advanced Energy Materials}\ } (\bibinfo
  {year} {2020}),\ 10.1002/aenm.201903242}\BibitemShut {NoStop}%
\bibitem [{\citenamefont {Mueller}\ \emph {et~al.}(2020)\citenamefont
  {Mueller}, \citenamefont {Hernandez},\ and\ \citenamefont
  {Wang}}]{Mueller2020}%
  \BibitemOpen
  \bibfield  {author} {\bibinfo {author} {\bibfnamefont {T.}~\bibnamefont
  {Mueller}}, \bibinfo {author} {\bibfnamefont {A.}~\bibnamefont {Hernandez}},
  \ and\ \bibinfo {author} {\bibfnamefont {C.}~\bibnamefont {Wang}},\ }\href
  {\doibase 10.1063/1.5126336} {\bibfield  {journal} {\bibinfo  {journal}
  {Journal of Chemical Physics}\ } (\bibinfo {year} {2020}),\
  10.1063/1.5126336}\BibitemShut {NoStop}%
\bibitem [{Sch(2019)}]{Schleder2019}%
  \BibitemOpen
  \href {\doibase 10.1088/2515-7639/ab084b} {\bibfield  {journal} {\bibinfo
  {journal} {Journal of Physics: Materials}\ } (\bibinfo {year} {2019}),\
  10.1088/2515-7639/ab084b}\BibitemShut {NoStop}%
\bibitem [{\citenamefont {Vandermause}\ \emph {et~al.}(2020)\citenamefont
  {Vandermause}, \citenamefont {Torrisi}, \citenamefont {Batzner},
  \citenamefont {Xie}, \citenamefont {Sun}, \citenamefont {Kolpak},\ and\
  \citenamefont {Kozinsky}}]{Vandermause2020}%
  \BibitemOpen
  \bibfield  {author} {\bibinfo {author} {\bibfnamefont {J.}~\bibnamefont
  {Vandermause}}, \bibinfo {author} {\bibfnamefont {S.~B.}\ \bibnamefont
  {Torrisi}}, \bibinfo {author} {\bibfnamefont {S.}~\bibnamefont {Batzner}},
  \bibinfo {author} {\bibfnamefont {Y.}~\bibnamefont {Xie}}, \bibinfo {author}
  {\bibfnamefont {L.}~\bibnamefont {Sun}}, \bibinfo {author} {\bibfnamefont
  {A.~M.}\ \bibnamefont {Kolpak}}, \ and\ \bibinfo {author} {\bibfnamefont
  {B.}~\bibnamefont {Kozinsky}},\ }\href {\doibase 10.1038/s41524-020-0283-z}
  {\bibfield  {journal} {\bibinfo  {journal} {npj Computational Materials}\ }
  (\bibinfo {year} {2020}),\ 10.1038/s41524-020-0283-z},\ \Eprint
  {http://arxiv.org/abs/1904.02042} {arXiv:1904.02042} \BibitemShut {NoStop}%
\bibitem [{\citenamefont {Jinnouchi}\ \emph {et~al.}(2019)\citenamefont
  {Jinnouchi}, \citenamefont {Lahnsteiner}, \citenamefont {Karsai},
  \citenamefont {Kresse},\ and\ \citenamefont {Bokdam}}]{Jinnouchi2019}%
  \BibitemOpen
  \bibfield  {author} {\bibinfo {author} {\bibfnamefont {R.}~\bibnamefont
  {Jinnouchi}}, \bibinfo {author} {\bibfnamefont {J.}~\bibnamefont
  {Lahnsteiner}}, \bibinfo {author} {\bibfnamefont {F.}~\bibnamefont {Karsai}},
  \bibinfo {author} {\bibfnamefont {G.}~\bibnamefont {Kresse}}, \ and\ \bibinfo
  {author} {\bibfnamefont {M.}~\bibnamefont {Bokdam}},\ }\href {\doibase
  10.1103/PhysRevLett.122.225701} {\bibfield  {journal} {\bibinfo  {journal}
  {Physical Review Letters}\ } (\bibinfo {year} {2019}),\
  10.1103/PhysRevLett.122.225701}\BibitemShut {NoStop}%
\bibitem [{\citenamefont {{Garrido Torres}}\ \emph {et~al.}(2019)\citenamefont
  {{Garrido Torres}}, \citenamefont {Jennings}, \citenamefont {Hansen},
  \citenamefont {Boes},\ and\ \citenamefont {Bligaard}}]{GarridoTorres2019}%
  \BibitemOpen
  \bibfield  {author} {\bibinfo {author} {\bibfnamefont {J.~A.}\ \bibnamefont
  {{Garrido Torres}}}, \bibinfo {author} {\bibfnamefont {P.~C.}\ \bibnamefont
  {Jennings}}, \bibinfo {author} {\bibfnamefont {M.~H.}\ \bibnamefont
  {Hansen}}, \bibinfo {author} {\bibfnamefont {J.~R.}\ \bibnamefont {Boes}}, \
  and\ \bibinfo {author} {\bibfnamefont {T.}~\bibnamefont {Bligaard}},\ }\href
  {\doibase 10.1103/PhysRevLett.122.156001} {\bibfield  {journal} {\bibinfo
  {journal} {Physical Review Letters}\ } (\bibinfo {year} {2019}),\
  10.1103/PhysRevLett.122.156001},\ \Eprint {http://arxiv.org/abs/1811.08022}
  {arXiv:1811.08022} \BibitemShut {NoStop}%
\bibitem [{\citenamefont {{Del R{\'{i}}o}}\ \emph {et~al.}(2019)\citenamefont
  {{Del R{\'{i}}o}}, \citenamefont {Mortensen},\ and\ \citenamefont
  {Jacobsen}}]{DelRio2019}%
  \BibitemOpen
  \bibfield  {author} {\bibinfo {author} {\bibfnamefont {E.~G.}\ \bibnamefont
  {{Del R{\'{i}}o}}}, \bibinfo {author} {\bibfnamefont {J.~J.}\ \bibnamefont
  {Mortensen}}, \ and\ \bibinfo {author} {\bibfnamefont {K.~W.}\ \bibnamefont
  {Jacobsen}},\ }\href {\doibase 10.1103/PhysRevB.100.104103} {\bibfield
  {journal} {\bibinfo  {journal} {Physical Review B}\ } (\bibinfo {year}
  {2019}),\ 10.1103/PhysRevB.100.104103},\ \Eprint
  {http://arxiv.org/abs/1808.08588} {arXiv:1808.08588} \BibitemShut {NoStop}%
\bibitem [{\citenamefont {Settles}(2010)}]{Settles2010}%
  \BibitemOpen
  \bibfield  {author} {\bibinfo {author} {\bibfnamefont {B.}~\bibnamefont
  {Settles}},\ }\href {\doibase 10.1.1.167.4245} {\bibfield  {journal}
  {\bibinfo  {journal} {Machine Learning}\ } (\bibinfo {year} {2010}),\
  10.1.1.167.4245}\BibitemShut {NoStop}%
\bibitem [{\citenamefont {Behler}\ and\ \citenamefont
  {Parrinello}(2007)}]{Behler2007}%
  \BibitemOpen
  \bibfield  {author} {\bibinfo {author} {\bibfnamefont {J.}~\bibnamefont
  {Behler}}\ and\ \bibinfo {author} {\bibfnamefont {M.}~\bibnamefont
  {Parrinello}},\ }\href {\doibase 10.1103/PhysRevLett.98.146401} {\bibfield
  {journal} {\bibinfo  {journal} {Physical Review Letters}\ } (\bibinfo {year}
  {2007}),\ 10.1103/PhysRevLett.98.146401}\BibitemShut {NoStop}%
\bibitem [{amp(2020)}]{amptorch}%
  \BibitemOpen
  \href@noop {} {} (\bibinfo {year} {2020}),\ \Eprint
  {http://arxiv.org/abs/https://github.com/ulissigroup/amptorch}
  {https://github.com/ulissigroup/amptorch} \BibitemShut {NoStop}%
\bibitem [{exa(2020)}]{examples}%
  \BibitemOpen
  \href@noop {} {} (\bibinfo {year} {2020}),\ \Eprint
  {http://arxiv.org/abs/https://github.com/ulissigroup/Enabling-Robust-Offline-Active-Learning-for-MLPs}
  {https://github.com/ulissigroup/Enabling-Robust-Offline-Active-Learning-for-MLPs}
  \BibitemShut {NoStop}%
\bibitem [{\citenamefont {Loshchilov}\ and\ \citenamefont
  {Hutter}(2019)}]{Loshchilov2019}%
  \BibitemOpen
  \bibfield  {author} {\bibinfo {author} {\bibfnamefont {I.}~\bibnamefont
  {Loshchilov}}\ and\ \bibinfo {author} {\bibfnamefont {F.}~\bibnamefont
  {Hutter}},\ }in\ \href@noop {} {\emph {\bibinfo {booktitle} {5th
  International Conference on Learning Representations, ICLR 2017 - Conference
  Track Proceedings}}}\ (\bibinfo {year} {2019})\ \Eprint
  {http://arxiv.org/abs/1608.03983} {arXiv:1608.03983} \BibitemShut {NoStop}%
\bibitem [{\citenamefont {Fey}\ and\ \citenamefont {Lenssen}(2019)}]{Fey2019}%
  \BibitemOpen
  \bibfield  {author} {\bibinfo {author} {\bibfnamefont {M.}~\bibnamefont
  {Fey}}\ and\ \bibinfo {author} {\bibfnamefont {J.~E.}\ \bibnamefont
  {Lenssen}},\ }\href {http://arxiv.org/abs/1903.02428} {\bibfield  {journal}
  {\bibinfo  {journal} {arXiv}\ } (\bibinfo {year} {2019})},\ \Eprint
  {http://arxiv.org/abs/1903.02428} {arXiv:1903.02428} \BibitemShut {NoStop}%
\bibitem [{\citenamefont {Paszke}\ \emph {et~al.}(2019)\citenamefont {Paszke},
  \citenamefont {Gross}, \citenamefont {Chintala}, \citenamefont {Chanan},
  \citenamefont {Yang}, \citenamefont {Facebook}, \citenamefont {Research},
  \citenamefont {Lin}, \citenamefont {Desmaison}, \citenamefont {Antiga},
  \citenamefont {Srl},\ and\ \citenamefont {Lerer}}]{Paszke2019}%
  \BibitemOpen
  \bibfield  {author} {\bibinfo {author} {\bibfnamefont {A.}~\bibnamefont
  {Paszke}}, \bibinfo {author} {\bibfnamefont {S.}~\bibnamefont {Gross}},
  \bibinfo {author} {\bibfnamefont {S.}~\bibnamefont {Chintala}}, \bibinfo
  {author} {\bibfnamefont {G.}~\bibnamefont {Chanan}}, \bibinfo {author}
  {\bibfnamefont {E.}~\bibnamefont {Yang}}, \bibinfo {author} {\bibfnamefont
  {Z.~D.}\ \bibnamefont {Facebook}}, \bibinfo {author} {\bibfnamefont {A.~I.}\
  \bibnamefont {Research}}, \bibinfo {author} {\bibfnamefont {Z.}~\bibnamefont
  {Lin}}, \bibinfo {author} {\bibfnamefont {A.}~\bibnamefont {Desmaison}},
  \bibinfo {author} {\bibfnamefont {L.}~\bibnamefont {Antiga}}, \bibinfo
  {author} {\bibfnamefont {O.}~\bibnamefont {Srl}}, \ and\ \bibinfo {author}
  {\bibfnamefont {A.}~\bibnamefont {Lerer}},\ }in\ \href@noop {} {\emph
  {\bibinfo {booktitle} {Advances in Neural Information Processing Systems
  32}}}\ (\bibinfo {year} {2019})\BibitemShut {NoStop}%
\bibitem [{\citenamefont {Bart{\~{o}}k}\ and\ \citenamefont
  {Cs{\'{a}}nyi}(2015)}]{Bartok2015}%
  \BibitemOpen
  \bibfield  {author} {\bibinfo {author} {\bibfnamefont {A.~P.}\ \bibnamefont
  {Bart{\~{o}}k}}\ and\ \bibinfo {author} {\bibfnamefont {G.}~\bibnamefont
  {Cs{\'{a}}nyi}},\ }\href {\doibase 10.1002/qua.24927} {\enquote {\bibinfo
  {title} {{Gaussian approximation potentials: A brief tutorial
  introduction}},}\ } (\bibinfo {year} {2015}),\ \Eprint
  {http://arxiv.org/abs/1502.01366} {arXiv:1502.01366} \BibitemShut {NoStop}%
\bibitem [{\citenamefont {Schran}\ \emph {et~al.}(2020)\citenamefont {Schran},
  \citenamefont {Behler},\ and\ \citenamefont {Marx}}]{Schran2020}%
  \BibitemOpen
  \bibfield  {author} {\bibinfo {author} {\bibfnamefont {C.}~\bibnamefont
  {Schran}}, \bibinfo {author} {\bibfnamefont {J.}~\bibnamefont {Behler}}, \
  and\ \bibinfo {author} {\bibfnamefont {D.}~\bibnamefont {Marx}},\ }\href
  {\doibase 10.1021/acs.jctc.9b00805} {\bibfield  {journal} {\bibinfo
  {journal} {Journal of Chemical Theory and Computation}\ } (\bibinfo {year}
  {2020}),\ 10.1021/acs.jctc.9b00805},\ \Eprint
  {http://arxiv.org/abs/1908.08734} {arXiv:1908.08734} \BibitemShut {NoStop}%
\bibitem [{\citenamefont {Willard}\ \emph {et~al.}()\citenamefont {Willard},
  \citenamefont {Jia}, \citenamefont {Xu}, \citenamefont {Steinbach},\ and\
  \citenamefont {Kumar}}]{Willard}%
  \BibitemOpen
  \bibfield  {author} {\bibinfo {author} {\bibfnamefont {J.~D.}\ \bibnamefont
  {Willard}}, \bibinfo {author} {\bibfnamefont {X.}~\bibnamefont {Jia}},
  \bibinfo {author} {\bibfnamefont {S.}~\bibnamefont {Xu}}, \bibinfo {author}
  {\bibfnamefont {M.}~\bibnamefont {Steinbach}}, \ and\ \bibinfo {author}
  {\bibfnamefont {V.}~\bibnamefont {Kumar}},\ }\href@noop {} {\emph {\bibinfo
  {title} {{Integrating Physics-Based Modeling with Machine Learning: A
  Survey}}}},\ \bibinfo {type} {Tech. Rep.},\ \Eprint
  {http://arxiv.org/abs/2003.04919v3} {arXiv:2003.04919v3} \BibitemShut
  {NoStop}%
\bibitem [{\citenamefont {Karpatne}\ \emph {et~al.}(2017)\citenamefont
  {Karpatne}, \citenamefont {Watkins}, \citenamefont {Read},\ and\
  \citenamefont {Kumar}}]{Karpatne2017}%
  \BibitemOpen
  \bibfield  {author} {\bibinfo {author} {\bibfnamefont {A.}~\bibnamefont
  {Karpatne}}, \bibinfo {author} {\bibfnamefont {W.}~\bibnamefont {Watkins}},
  \bibinfo {author} {\bibfnamefont {J.}~\bibnamefont {Read}}, \ and\ \bibinfo
  {author} {\bibfnamefont {V.}~\bibnamefont {Kumar}},\ }\href@noop {}
  {\bibfield  {journal} {\bibinfo  {journal} {arXiv preprint arXiv:1710.11431}\
  } (\bibinfo {year} {2017})}\BibitemShut {NoStop}%
\bibitem [{\citenamefont {Ramakrishnan}\ \emph {et~al.}(2014)\citenamefont
  {Ramakrishnan}, \citenamefont {Dral}, \citenamefont {Rupp},\ and\
  \citenamefont {{Von Lilienfeld}}}]{Ramakrishnan2014}%
  \BibitemOpen
  \bibfield  {author} {\bibinfo {author} {\bibfnamefont {R.}~\bibnamefont
  {Ramakrishnan}}, \bibinfo {author} {\bibfnamefont {P.~O.}\ \bibnamefont
  {Dral}}, \bibinfo {author} {\bibfnamefont {M.}~\bibnamefont {Rupp}}, \ and\
  \bibinfo {author} {\bibfnamefont {O.~A.}\ \bibnamefont {{Von Lilienfeld}}},\
  }\href {\doibase 10.1038/sdata.2014.22} {\bibfield  {journal} {\bibinfo
  {journal} {Scientific Data}\ } (\bibinfo {year} {2014}),\
  10.1038/sdata.2014.22}\BibitemShut {NoStop}%
\bibitem [{\citenamefont {Zhu}\ \emph {et~al.}(2019)\citenamefont {Zhu},
  \citenamefont {Vuong}, \citenamefont {Sumpter},\ and\ \citenamefont
  {Irle}}]{Zhu2019}%
  \BibitemOpen
  \bibfield  {author} {\bibinfo {author} {\bibfnamefont {J.}~\bibnamefont
  {Zhu}}, \bibinfo {author} {\bibfnamefont {V.~Q.}\ \bibnamefont {Vuong}},
  \bibinfo {author} {\bibfnamefont {B.~G.}\ \bibnamefont {Sumpter}}, \ and\
  \bibinfo {author} {\bibfnamefont {S.}~\bibnamefont {Irle}},\ }\href {\doibase
  10.1557/mrc.2019.80} {\bibfield  {journal} {\bibinfo  {journal} {MRS
  Communications}\ } (\bibinfo {year} {2019}),\
  10.1557/mrc.2019.80}\BibitemShut {NoStop}%
\bibitem [{\citenamefont {Loshchilov}\ and\ \citenamefont
  {Hutter}(2016)}]{loshchilov2016sgdr}%
  \BibitemOpen
  \bibfield  {author} {\bibinfo {author} {\bibfnamefont {I.}~\bibnamefont
  {Loshchilov}}\ and\ \bibinfo {author} {\bibfnamefont {F.}~\bibnamefont
  {Hutter}},\ }\href@noop {} {\enquote {\bibinfo {title} {Sgdr: Stochastic
  gradient descent with warm restarts},}\ } (\bibinfo {year} {2016}),\ \Eprint
  {http://arxiv.org/abs/1608.03983} {arXiv:1608.03983 [cs.LG]} \BibitemShut
  {NoStop}%
\bibitem [{\citenamefont {Peterson}\ \emph {et~al.}(2017)\citenamefont
  {Peterson}, \citenamefont {Christensen},\ and\ \citenamefont
  {Khorshidi}}]{Peterson2017}%
  \BibitemOpen
  \bibfield  {author} {\bibinfo {author} {\bibfnamefont {A.~A.}\ \bibnamefont
  {Peterson}}, \bibinfo {author} {\bibfnamefont {R.}~\bibnamefont
  {Christensen}}, \ and\ \bibinfo {author} {\bibfnamefont {A.}~\bibnamefont
  {Khorshidi}},\ }\href {\doibase 10.1039/c7cp00375g} {\bibfield  {journal}
  {\bibinfo  {journal} {Physical Chemistry Chemical Physics}\ } (\bibinfo
  {year} {2017}),\ 10.1039/c7cp00375g}\BibitemShut {NoStop}%
\bibitem [{\citenamefont {Tran}\ \emph {et~al.}(2020)\citenamefont {Tran},
  \citenamefont {Neiswanger}, \citenamefont {Yoon}, \citenamefont {Zhang},
  \citenamefont {Xing},\ and\ \citenamefont {Ulissi}}]{Tran2020}%
  \BibitemOpen
  \bibfield  {author} {\bibinfo {author} {\bibfnamefont {K.}~\bibnamefont
  {Tran}}, \bibinfo {author} {\bibfnamefont {W.}~\bibnamefont {Neiswanger}},
  \bibinfo {author} {\bibfnamefont {J.}~\bibnamefont {Yoon}}, \bibinfo {author}
  {\bibfnamefont {Q.}~\bibnamefont {Zhang}}, \bibinfo {author} {\bibfnamefont
  {E.}~\bibnamefont {Xing}}, \ and\ \bibinfo {author} {\bibfnamefont {Z.~W.}\
  \bibnamefont {Ulissi}},\ }\href {\doibase 10.1088/2632-2153/ab7e1a}
  {\bibfield  {journal} {\bibinfo  {journal} {Machine Learning: Science and
  Technology}\ } (\bibinfo {year} {2020}),\ 10.1088/2632-2153/ab7e1a},\ \Eprint
  {http://arxiv.org/abs/1912.10066} {arXiv:1912.10066} \BibitemShut {NoStop}%
\bibitem [{\citenamefont {Kresse}\ and\ \citenamefont
  {Hafner}(1993)}]{Kresse1993}%
  \BibitemOpen
  \bibfield  {author} {\bibinfo {author} {\bibfnamefont {G.}~\bibnamefont
  {Kresse}}\ and\ \bibinfo {author} {\bibfnamefont {J.}~\bibnamefont
  {Hafner}},\ }\href {\doibase 10.1103/PhysRevB.48.13115} {\bibfield  {journal}
  {\bibinfo  {journal} {Physical Review B}\ } (\bibinfo {year} {1993}),\
  10.1103/PhysRevB.48.13115}\BibitemShut {NoStop}%
\bibitem [{\citenamefont {Kresse}\ and\ \citenamefont
  {Furthm{\"{u}}ller}(1996)}]{Kresse1996}%
  \BibitemOpen
  \bibfield  {author} {\bibinfo {author} {\bibfnamefont {G.}~\bibnamefont
  {Kresse}}\ and\ \bibinfo {author} {\bibfnamefont {J.}~\bibnamefont
  {Furthm{\"{u}}ller}},\ }\href {\doibase 10.1016/0927-0256(96)00008-0}
  {\bibfield  {journal} {\bibinfo  {journal} {Computational Materials Science}\
  } (\bibinfo {year} {1996}),\ 10.1016/0927-0256(96)00008-0}\BibitemShut
  {NoStop}%
\bibitem [{\citenamefont {Giannozzi}\ \emph {et~al.}(2009)\citenamefont
  {Giannozzi}, \citenamefont {Baroni}, \citenamefont {Bonini}, \citenamefont
  {Calandra}, \citenamefont {Car}, \citenamefont {Cavazzoni}, \citenamefont
  {Ceresoli}, \citenamefont {Chiarotti}, \citenamefont {Cococcioni},
  \citenamefont {Dabo}, \citenamefont {{Dal Corso}}, \citenamefont {{De
  Gironcoli}}, \citenamefont {Fabris}, \citenamefont {Fratesi}, \citenamefont
  {Gebauer}, \citenamefont {Gerstmann}, \citenamefont {Gougoussis},
  \citenamefont {Kokalj}, \citenamefont {Lazzeri}, \citenamefont
  {Martin-Samos}, \citenamefont {Marzari}, \citenamefont {Mauri}, \citenamefont
  {Mazzarello}, \citenamefont {Paolini}, \citenamefont {Pasquarello},
  \citenamefont {Paulatto}, \citenamefont {Sbraccia}, \citenamefont {Scandolo},
  \citenamefont {Sclauzero}, \citenamefont {Seitsonen}, \citenamefont
  {Smogunov}, \citenamefont {Umari},\ and\ \citenamefont
  {Wentzcovitch}}]{Giannozzi2009}%
  \BibitemOpen
  \bibfield  {author} {\bibinfo {author} {\bibfnamefont {P.}~\bibnamefont
  {Giannozzi}}, \bibinfo {author} {\bibfnamefont {S.}~\bibnamefont {Baroni}},
  \bibinfo {author} {\bibfnamefont {N.}~\bibnamefont {Bonini}}, \bibinfo
  {author} {\bibfnamefont {M.}~\bibnamefont {Calandra}}, \bibinfo {author}
  {\bibfnamefont {R.}~\bibnamefont {Car}}, \bibinfo {author} {\bibfnamefont
  {C.}~\bibnamefont {Cavazzoni}}, \bibinfo {author} {\bibfnamefont
  {D.}~\bibnamefont {Ceresoli}}, \bibinfo {author} {\bibfnamefont {G.~L.}\
  \bibnamefont {Chiarotti}}, \bibinfo {author} {\bibfnamefont {M.}~\bibnamefont
  {Cococcioni}}, \bibinfo {author} {\bibfnamefont {I.}~\bibnamefont {Dabo}},
  \bibinfo {author} {\bibfnamefont {A.}~\bibnamefont {{Dal Corso}}}, \bibinfo
  {author} {\bibfnamefont {S.}~\bibnamefont {{De Gironcoli}}}, \bibinfo
  {author} {\bibfnamefont {S.}~\bibnamefont {Fabris}}, \bibinfo {author}
  {\bibfnamefont {G.}~\bibnamefont {Fratesi}}, \bibinfo {author} {\bibfnamefont
  {R.}~\bibnamefont {Gebauer}}, \bibinfo {author} {\bibfnamefont
  {U.}~\bibnamefont {Gerstmann}}, \bibinfo {author} {\bibfnamefont
  {C.}~\bibnamefont {Gougoussis}}, \bibinfo {author} {\bibfnamefont
  {A.}~\bibnamefont {Kokalj}}, \bibinfo {author} {\bibfnamefont
  {M.}~\bibnamefont {Lazzeri}}, \bibinfo {author} {\bibfnamefont
  {L.}~\bibnamefont {Martin-Samos}}, \bibinfo {author} {\bibfnamefont
  {N.}~\bibnamefont {Marzari}}, \bibinfo {author} {\bibfnamefont
  {F.}~\bibnamefont {Mauri}}, \bibinfo {author} {\bibfnamefont
  {R.}~\bibnamefont {Mazzarello}}, \bibinfo {author} {\bibfnamefont
  {S.}~\bibnamefont {Paolini}}, \bibinfo {author} {\bibfnamefont
  {A.}~\bibnamefont {Pasquarello}}, \bibinfo {author} {\bibfnamefont
  {L.}~\bibnamefont {Paulatto}}, \bibinfo {author} {\bibfnamefont
  {C.}~\bibnamefont {Sbraccia}}, \bibinfo {author} {\bibfnamefont
  {S.}~\bibnamefont {Scandolo}}, \bibinfo {author} {\bibfnamefont
  {G.}~\bibnamefont {Sclauzero}}, \bibinfo {author} {\bibfnamefont {A.~P.}\
  \bibnamefont {Seitsonen}}, \bibinfo {author} {\bibfnamefont {A.}~\bibnamefont
  {Smogunov}}, \bibinfo {author} {\bibfnamefont {P.}~\bibnamefont {Umari}}, \
  and\ \bibinfo {author} {\bibfnamefont {R.~M.}\ \bibnamefont {Wentzcovitch}},\
  }\href {\doibase 10.1088/0953-8984/21/39/395502} {\bibfield  {journal}
  {\bibinfo  {journal} {Journal of Physics Condensed Matter}\ } (\bibinfo
  {year} {2009}),\ 10.1088/0953-8984/21/39/395502},\ \Eprint
  {http://arxiv.org/abs/0906.2569} {arXiv:0906.2569} \BibitemShut {NoStop}%
\bibitem [{\citenamefont {Henkelman}\ \emph {et~al.}(2000)\citenamefont
  {Henkelman}, \citenamefont {Uberuaga},\ and\ \citenamefont
  {J{\'{o}}nsson}}]{Henkelman2000}%
  \BibitemOpen
  \bibfield  {author} {\bibinfo {author} {\bibfnamefont {G.}~\bibnamefont
  {Henkelman}}, \bibinfo {author} {\bibfnamefont {B.~P.}\ \bibnamefont
  {Uberuaga}}, \ and\ \bibinfo {author} {\bibfnamefont {H.}~\bibnamefont
  {J{\'{o}}nsson}},\ }\href {\doibase 10.1063/1.1329672} {\bibfield  {journal}
  {\bibinfo  {journal} {Journal of Chemical Physics}\ } (\bibinfo {year}
  {2000}),\ 10.1063/1.1329672}\BibitemShut {NoStop}%
\bibitem [{\citenamefont {Henkelman}\ and\ \citenamefont
  {J{\'{o}}nsson}(2000)}]{Henkelman2000_2}%
  \BibitemOpen
  \bibfield  {author} {\bibinfo {author} {\bibfnamefont {G.}~\bibnamefont
  {Henkelman}}\ and\ \bibinfo {author} {\bibfnamefont {H.}~\bibnamefont
  {J{\'{o}}nsson}},\ }\href {\doibase 10.1063/1.1323224} {\bibfield  {journal}
  {\bibinfo  {journal} {Journal of Chemical Physics}\ } (\bibinfo {year}
  {2000}),\ 10.1063/1.1323224}\BibitemShut {NoStop}%
\bibitem [{\citenamefont {Ang}\ \emph {et~al.}(2020)\citenamefont {Ang},
  \citenamefont {Wang}, \citenamefont {Schwalbe-Koda}, \citenamefont
  {Axelrod},\ and\ \citenamefont {Gomez-Bombarelli}}]{Ang2020}%
  \BibitemOpen
  \bibfield  {author} {\bibinfo {author} {\bibfnamefont {S.~J.}\ \bibnamefont
  {Ang}}, \bibinfo {author} {\bibfnamefont {W.}~\bibnamefont {Wang}}, \bibinfo
  {author} {\bibfnamefont {D.}~\bibnamefont {Schwalbe-Koda}}, \bibinfo {author}
  {\bibfnamefont {S.}~\bibnamefont {Axelrod}}, \ and\ \bibinfo {author}
  {\bibfnamefont {R.}~\bibnamefont {Gomez-Bombarelli}},\ }\href {\doibase
  10.26434/CHEMRXIV.11910948.V2} {\bibfield  {journal} {\bibinfo  {journal}
  {ChemRxiv}\ } (\bibinfo {year} {2020}),\
  10.26434/CHEMRXIV.11910948.V2}\BibitemShut {NoStop}%
\bibitem [{\citenamefont {{Hjorth Larsen}}\ \emph {et~al.}(2017)\citenamefont
  {{Hjorth Larsen}}, \citenamefont {{J{\O}rgen Mortensen}}, \citenamefont
  {Blomqvist}, \citenamefont {Castelli}, \citenamefont {Christensen},
  \citenamefont {Du{\l}ak}, \citenamefont {Friis}, \citenamefont {Groves},
  \citenamefont {Hammer}, \citenamefont {Hargus}, \citenamefont {Hermes},
  \citenamefont {Jennings}, \citenamefont {{Bjerre Jensen}}, \citenamefont
  {Kermode}, \citenamefont {Kitchin}, \citenamefont {{Leonhard Kolsbjerg}},
  \citenamefont {Kubal}, \citenamefont {Kaasbjerg}, \citenamefont {Lysgaard},
  \citenamefont {{Bergmann Maronsson}}, \citenamefont {Maxson}, \citenamefont
  {Olsen}, \citenamefont {Pastewka}, \citenamefont {Peterson}, \citenamefont
  {Rostgaard}, \citenamefont {Schi{\O}tz}, \citenamefont {Sch{\"{u}}tt},
  \citenamefont {Strange}, \citenamefont {Thygesen}, \citenamefont {Vegge},
  \citenamefont {Vilhelmsen}, \citenamefont {Walter}, \citenamefont {Zeng},\
  and\ \citenamefont {Jacobsen}}]{HjorthLarsen2017}%
  \BibitemOpen
  \bibfield  {author} {\bibinfo {author} {\bibfnamefont {A.}~\bibnamefont
  {{Hjorth Larsen}}}, \bibinfo {author} {\bibfnamefont {J.}~\bibnamefont
  {{J{\O}rgen Mortensen}}}, \bibinfo {author} {\bibfnamefont {J.}~\bibnamefont
  {Blomqvist}}, \bibinfo {author} {\bibfnamefont {I.~E.}\ \bibnamefont
  {Castelli}}, \bibinfo {author} {\bibfnamefont {R.}~\bibnamefont
  {Christensen}}, \bibinfo {author} {\bibfnamefont {M.}~\bibnamefont
  {Du{\l}ak}}, \bibinfo {author} {\bibfnamefont {J.}~\bibnamefont {Friis}},
  \bibinfo {author} {\bibfnamefont {M.~N.}\ \bibnamefont {Groves}}, \bibinfo
  {author} {\bibfnamefont {B.}~\bibnamefont {Hammer}}, \bibinfo {author}
  {\bibfnamefont {C.}~\bibnamefont {Hargus}}, \bibinfo {author} {\bibfnamefont
  {E.~D.}\ \bibnamefont {Hermes}}, \bibinfo {author} {\bibfnamefont {P.~C.}\
  \bibnamefont {Jennings}}, \bibinfo {author} {\bibfnamefont {P.}~\bibnamefont
  {{Bjerre Jensen}}}, \bibinfo {author} {\bibfnamefont {J.}~\bibnamefont
  {Kermode}}, \bibinfo {author} {\bibfnamefont {J.~R.}\ \bibnamefont
  {Kitchin}}, \bibinfo {author} {\bibfnamefont {E.}~\bibnamefont {{Leonhard
  Kolsbjerg}}}, \bibinfo {author} {\bibfnamefont {J.}~\bibnamefont {Kubal}},
  \bibinfo {author} {\bibfnamefont {K.}~\bibnamefont {Kaasbjerg}}, \bibinfo
  {author} {\bibfnamefont {S.}~\bibnamefont {Lysgaard}}, \bibinfo {author}
  {\bibfnamefont {J.}~\bibnamefont {{Bergmann Maronsson}}}, \bibinfo {author}
  {\bibfnamefont {T.}~\bibnamefont {Maxson}}, \bibinfo {author} {\bibfnamefont
  {T.}~\bibnamefont {Olsen}}, \bibinfo {author} {\bibfnamefont
  {L.}~\bibnamefont {Pastewka}}, \bibinfo {author} {\bibfnamefont
  {A.}~\bibnamefont {Peterson}}, \bibinfo {author} {\bibfnamefont
  {C.}~\bibnamefont {Rostgaard}}, \bibinfo {author} {\bibfnamefont
  {J.}~\bibnamefont {Schi{\O}tz}}, \bibinfo {author} {\bibfnamefont
  {O.}~\bibnamefont {Sch{\"{u}}tt}}, \bibinfo {author} {\bibfnamefont
  {M.}~\bibnamefont {Strange}}, \bibinfo {author} {\bibfnamefont {K.~S.}\
  \bibnamefont {Thygesen}}, \bibinfo {author} {\bibfnamefont {T.}~\bibnamefont
  {Vegge}}, \bibinfo {author} {\bibfnamefont {L.}~\bibnamefont {Vilhelmsen}},
  \bibinfo {author} {\bibfnamefont {M.}~\bibnamefont {Walter}}, \bibinfo
  {author} {\bibfnamefont {Z.}~\bibnamefont {Zeng}}, \ and\ \bibinfo {author}
  {\bibfnamefont {K.~W.}\ \bibnamefont {Jacobsen}},\ }\href {\doibase
  10.1088/1361-648X/aa680e} {\bibfield  {journal} {\bibinfo  {journal} {Journal
  of Physics Condensed Matter}\ } (\bibinfo {year} {2017}),\
  10.1088/1361-648X/aa680e}\BibitemShut {NoStop}%
\bibitem [{\citenamefont {Perdew}\ \emph {et~al.}(1996)\citenamefont {Perdew},
  \citenamefont {Burke},\ and\ \citenamefont {Ernzerhof}}]{Perdew1996}%
  \BibitemOpen
  \bibfield  {author} {\bibinfo {author} {\bibfnamefont {J.~P.}\ \bibnamefont
  {Perdew}}, \bibinfo {author} {\bibfnamefont {K.}~\bibnamefont {Burke}}, \
  and\ \bibinfo {author} {\bibfnamefont {M.}~\bibnamefont {Ernzerhof}},\ }\href
  {\doibase 10.1103/PhysRevLett.77.3865} {\bibfield  {journal} {\bibinfo
  {journal} {Physical Review Letters}\ } (\bibinfo {year} {1996}),\
  10.1103/PhysRevLett.77.3865}\BibitemShut {NoStop}%
\bibitem [{\citenamefont {Garrity}\ \emph {et~al.}(2014)\citenamefont
  {Garrity}, \citenamefont {Bennett}, \citenamefont {Rabe},\ and\ \citenamefont
  {Vanderbilt}}]{Garrity2014}%
  \BibitemOpen
  \bibfield  {author} {\bibinfo {author} {\bibfnamefont {K.~F.}\ \bibnamefont
  {Garrity}}, \bibinfo {author} {\bibfnamefont {J.~W.}\ \bibnamefont
  {Bennett}}, \bibinfo {author} {\bibfnamefont {K.~M.}\ \bibnamefont {Rabe}}, \
  and\ \bibinfo {author} {\bibfnamefont {D.}~\bibnamefont {Vanderbilt}},\
  }\href {\doibase 10.1016/j.commatsci.2013.08.053} {\bibfield  {journal}
  {\bibinfo  {journal} {Computational Materials Science}\ } (\bibinfo {year}
  {2014}),\ 10.1016/j.commatsci.2013.08.053},\ \Eprint
  {http://arxiv.org/abs/1305.5973} {arXiv:1305.5973} \BibitemShut {NoStop}%
\end{thebibliography}%
\clearpage

\end{document}